# Effects of degumming conditions on the structure of the regenerated silk fibroin and the properties of its film

Ruixue Sun[a]    Junli Hu*,[a]

**Abstract:** The traditional degumming method using sodium carbonate solution severely damages the structure of silk fibroin and results in low molecular weight, which limits the properties and applications of silk materials. In this study, we report a modified degumming method and compared it with the traditional one. The results indicate that compared with the traditional degumming method, the modified degumming method produces silk fibroin with higher β-sheet content (53.43% vs 46.68%), higher crystallinity (44.92% vs 42.72%), and more concentrated molecular weight (100 ~ 300 kDa vs 72 ~ 300 kDa), while allowing for the same full removal of sericin. Correspondingly, the film of the silk fibroin prepared with the modified degumming method exhibits higher β-sheet content (35.22% vs 32.82%), higher crystallinity (26.96% vs 25.53%), and higher tensile strength (87.91 MPa vs 72.62 MPa). This work provides a degumming method superior to the traditional one and lays a foundation for the development of high-performance silk materials.

## 1. Introduction

Silk fibroin is the main component of raw silk, has a variable biodegradability and adjustable biocompatibility, and is simple to into a variety of forms including hydrogels[1-3], fibers[4], films[5,6], microspheres[7], tube scaffolds[8-10], sponge scaffolds[11]. In particular, silk films hold significant potential for applications in the fields of biomedicine[12,13], food preservation[14-16], biosensing[17], and environmental protection[18,19]. The regeneration of silk fibroin from raw silk include steps of degumming, dissolution, dialysis and centrifugation, among which the degumming step has the greatest impact on the structure of the regenerated silk fibroin and prepared silk materials. The traditional degumming method involves boiling raw silk with 0.05% sodium carbonate solution at a bath ratio of 1:100 for 1.5 hours. Although the use of sodium carbonate solution excessively at high temperature in this method can remove all sericin, studies have consistently shown that it also damages the structure of the silk fibroin [20,21].

In this work, we report a modified degumming method different from the traditional one, that is boiling raw silk with 0.05% sodium carbonate solution at a bath ratio of 1:100 for 0.5 hours, followed by boiling raw silk with pure water at a bath ratio of 1:100 for 1 hour. We found that the modified degumming method can fully remove sericin while less harm silk fibroin. The regenerated silk fibroin from the modified degumming method has higher β-sheet content, crystallinity, and more concentrated molecular weight than that from the traditional degumming method, and correspondingly, the film of the silk fibroin prepared with the modified degumming method exhibits higher β-sheet content, crystallinity, and tensile strength. This work provides a degumming method superior to the traditional one and lays a foundation for the development of high-performance silk materials.

---

[a] Key Laboratory of UV-Emitting Materials and Technology, Ministry of Education, Northeast Normal University, Changchun 130024, China

* E-mail: hujl100@nenu.edu.cn

## 2. Materials and methods

### 2.1 Preparation of the silk films

#### 2.2.1 Silk degumming

Traditional degumming method: The silk cocoons were degummed by boiling thrice in 0.05% $Na_2CO_3$ solution for 30 min to extract sericin, then thoroughly washed with distilled water and dried at 60 °C.

Modified degumming method: Cocoons were boiled for 30 min in an aqueous solution of 0.05% $Na_2CO_3$ and 30 min in boiled water twice, rinsed thoroughly with distilled water and dried at 60 °C.

#### 2.2.2 Preparation of silk fibroin films

Two kinds of silk fibers were dissolved in 9.3 M LiBr solution at 60 °C for 4 h. Then the fibroin solutions were filtered and dialyzed against distilled water for 2 days. Then we put them into a high-speed centrifuge and centrifuged at 9000 rpm and 20 min to remove insoluble impurities and undissolved parts of the silk to obtain the regenerated silk fibroin solution. Dilute the silk fibroin solution to a mass fraction of 5%. The films were prepared by casting and then treated with water vapor at 40 °C for 0.5 h.

### 2.3 Characterizations

#### 2.3.1 SEM imaging

After gold was sprayed on the surface of the degummed silk by ion sputtering instrument for 90 s, the surface morphology of the silk was observed by scanning electron microscope (20 kV). The silk fiber sizes were determined using the Image J.

#### 2.3.2 Degumming rate

The degumming rate of silk was characterized by drying weighing method. Weigh the dried crushed silkworm shells and weigh their mass. Then the degummed silk is dried at room temperature until the remains unchanged, and the degummed rate of the silk is calculated according to its mass.

#### 2.3.3 FTIR analysis

The sample of 4 mg silk and 400 mg KBr powder were fully mixed and then pressed into presses. The scanning results in the range of 4000 ~ 600 $cm^{-1}$ were recorded by FTIR with a resolution of 4 $cm^{-1}$. Peak Fit was used to analysis 1700 ~ 1600 $cm^{-1}$ with Gauss-Lorentz composite function, and the secondary structure contents of each silk were calculated.

#### 2.3.4 Crystallization characterization

The X-ray has a wavelength of 0.1541 nm, a scanning speed of 10 °/min, and a scanning range of

5 to 50 °. Using Origin, the obtained XRD spectra were fitted with Gaussian to obtain the area of each diffraction peak and calculate the crystallinity of the silk, that is, the ratio of the sum of the peak area of the crystallization diffraction peak to the total area.

2.3.5 Molecular weight test

The separation gel of 6%、10% and the concentration gel of 5% were prepared, until the gel was fully polymerized. During electrophoresis, the voltage in the concentrated gel zone was 80 V, and the voltage in the separation gel zone to 120 V. After adding Coomassie brilliant blue, decolorizing reagent was used to decolorize the gel background until it became transparent, and record with phone.

2.3.6 Mechanical properties

The mechanical properties of fibroin films were tested by a universal material testing machine at room temperature. The samples were cut into 20 mm×6 mm pieces and tested at 1 mm/min. When the mechanical properties of silk fibroin films were measured in the wet state, they were soaked in phosphate buffer solution until completely swollen. Three samples were tested in parallel.

2.3.7 Vapor permeability rate

The silk films were cut into 30 mm×30 mm squares and cover centrifuge tubes with 20 mL solution (0.142 mol/L NaCl and 0.025 mol/L $CaCl_2$). The whole device was weighed $W_1$. Cultured for 24 h at 37 °C and weighed again ($W_2$). The vapor permeability rate of the films was calculated as $(W_1-W_2)/A$. Where A is the cross-sectional area of the centrifugal tube mouth. Each sample has at least 3 parallel samples.

2.3.8 Contact angle

Cut the silk films into rectangles and paste them on a plastic slide. Add 2 μL of ultra-pure water to the surface of the films. The dynamic process that silk films absorb solutions were observed by shooting video to study their surface properties and wettability.

2.3.9 Swelling rate

The mass of the 2 cm× 2 cm silk fibroin film was measured as $N_1$. It was fully soaked in 0.01 M PBS buffer, and the silk fibroin films were taken out every 0.5 h and the water on the surface of the films were carefully absorbed by filter paper, then weighed again and recorded as $N_2$, and the swelling rate of the sample $Q= (N_2–N_1)/N_1×100\%$ was calculated.

2.3.10 Degradation rate

The mass $P_1$ of the silk fibroin films were weighed and put them into the bottle and completely immersed in 5 mL 0.01m PBS buffer or 0.01M PBS buffer with 100 μg/mL and 15 μg/mL protease K, and then the bottles were cultured at 37 °C. Take out the films at regular intervals and put them in the oven at 60 °C for drying and weighed again ($P_2$). The degradation rate $C= (P_2–P_1)/P_1$ was calculated according to the formula.

# 3 Results and discussion

## 3.1 Degummed silk fibers

Fig. 1a shows the SEM images of the two kinds of degummed silk. The surfaces of the degummed silk made using the conventional degumming method had grooves on them. It was evident that some of the silk fibers had peeled off, indicating significant damage. Conversely, the smooth surfaces have nearly no rough bumps on the fibers degummed with the modified method, indicating that the modified degumming method was superior to the traditional one in that it could remove sericin the same cleanly but more gently.

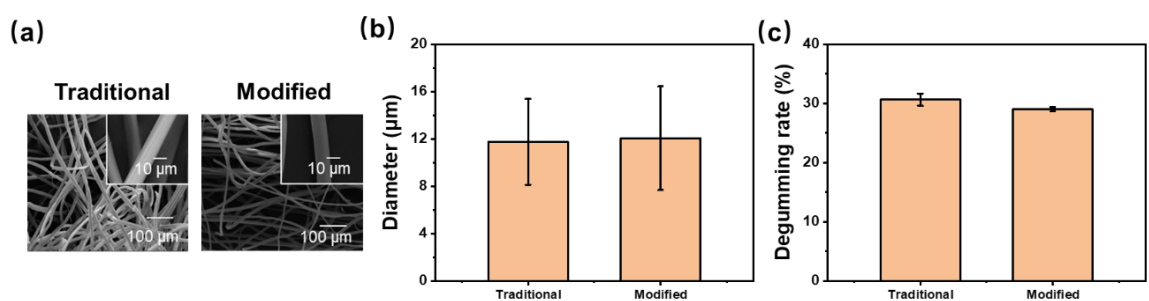

Fig.1 (a) SEM images, (b)Diameter, (c)Degumming rate.

As shown in Fig. 1b, according to the diameter statistical analysis of the two types of degummed silk, the diameter of the silk obtained using modified method is slightly larger than that using the traditional method was (12.06 μm vs 11.74 μm). In the meanwhile, the degumming rates of the modified method was slightly lower than the traditional method (29.01% vs 30.63%). These results further demonstrated that the modified method did less damage to silk fibroin.

The structure of the traditional silk were the same as that of the modified silk (Fig. 1c). Their peak both are at 1656 $cm^{-1}$ (amide I band) corresponding to the α-helix structure and 1518 $cm^{-1}$, which was associated with the β-sheet. In addition, their characteristic peaks are at 1227 $cm^{-1}$ representing the structure of Silk I, 1162 $cm^{-1}$ related to N-Cα chemical bond , 997 $cm^{-1}$ and 975 $cm^{-1}$ by silk fibroin "glycine (Gly) -Gly" and "Gly-alanine (Ala)".[22-24]. The principal β-sheet structure was present in two types of degummed silk. We analyzed the peptide skeleton of amide I in the infrared spectral region of 1700 ~ 1600 $cm^{-1}$ to examine various secondary structures in fibroin proteins. The β-sheet content of traditional degummed silk was 46.68%, which was lower than that of modified degummed silk (53.43%). The result showed that the modified silk fibers were more complete and not degraded significantly, which were superior to the traditional fibers.

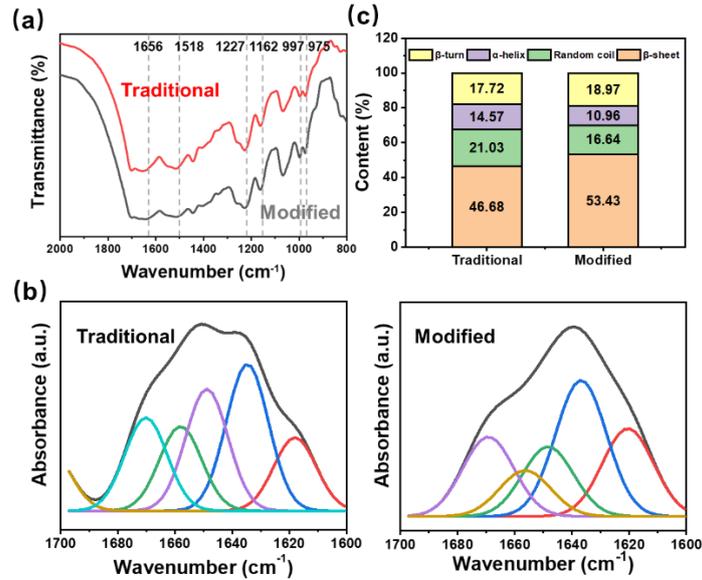

Fig.2 (a) FTIR spectra, (b) FTIR Curve-fitting for FSD amide I spectra, (c) Secondary structure content.

There were two primary crystal types in the two types of degummed silk: Silk I and Silk II(Fig. 3a). The crystallinity of the traditional silk is 42.72%, while the modified silk fibroin has higher crystallinity(42.92%)(Fig. 3c). This can also prove that compared with traditional silk fibroin fibers, modified silk fibroin fibers are less degraded and destroyed.

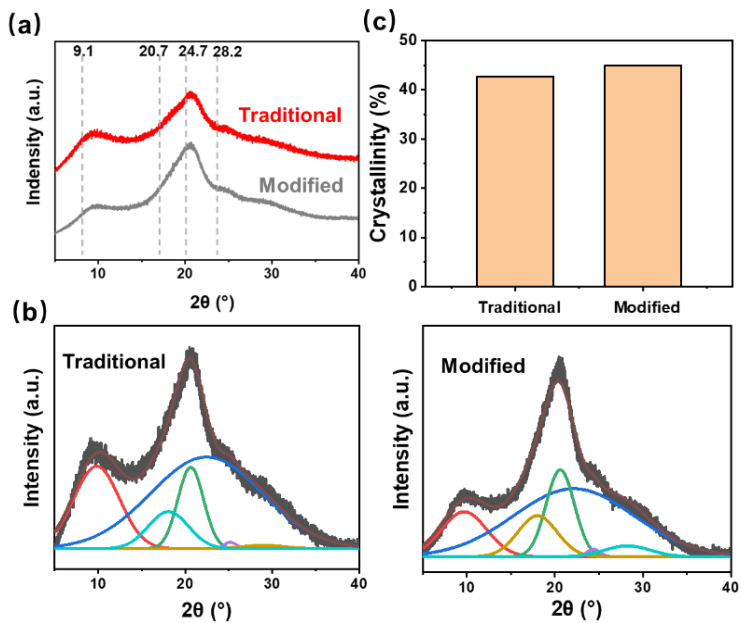

Fig.3 (a) XRD curves, (b) XRD pattern peak fitting diagrams, (c)Crystallinity.

The traditional silk had a molecular weight primarily distributed between 72 ~ 300 kDa as shown by SDS-PAGE pictures (Fig. 4), whereas the modified silk fibroin molecular weight is about 100 ~ 300 kDa approximately. The modified silk fibroin fibers had a relatively larger molecular weight than the traditional fibers, it is also confirmed that the improved degumming scheme can reduce the destruction of silk fibroin protein.

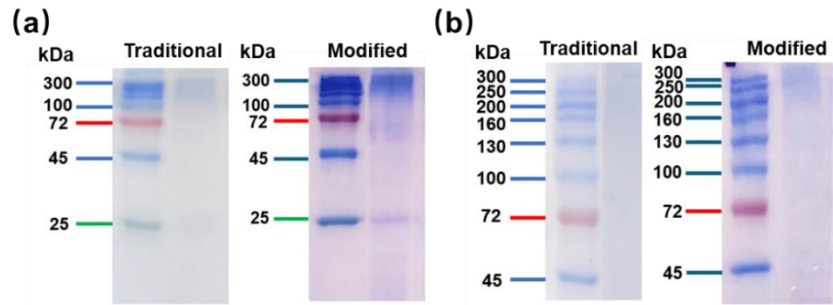

Fig.4 SDS-PAGE electrophoretic picture (a)The separation gel of 10%, (b) The separation gel of 6%.

In summary, the modified degumming scheme combined 0.5 hours of 0.05% sodium carbonate solution with 1.0 hours of pure water was superior to the traditional 1.5 hours method in that it removed sericin more efficiently, caused less damage to silk fibroin and retained a higher molecular weight.

3.2 Silk films

3.2.1 Molecular conformation

There are no discernible differences between the traditional and modified degummed silk films in Fourier transform infrared spectra (Fig. 5a). Their β-sheet structures were associated with the peaks at 1625 cm$^{-1}$ and 1515 cm$^{-1}$, while the random coil was linked to the peak at 1240 cm$^{-1}$[25,26]. SF-T and SF-M had β-sheet contents of 32.82% and 35.22% as well (Fig. 5c). Compared with the traditional films, the modified films had more β-sheet structure, because the improved degumming fibers retain a higher molecular weight of silk fibroin and have more β-sheet structure. The resulting SF-M films have longer molecular chains resulting in stronger intermolecular interactions and higher β-sheets.

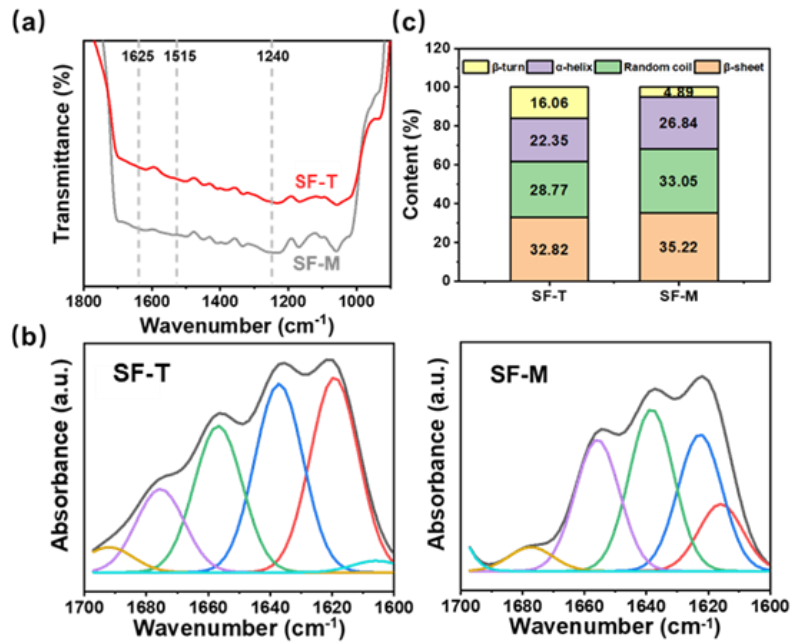

Fig.5 (a) FTIR spectra, (b) FTIR Curve-fitting for FSD amide I spectra, (c) Secondary structure content.

3.2.2 Crystallization properties

The XRD curves of silk films (Fig. 6a) were mainly consisted of Silk I and Silk II[27]. Two types of films peaks both at 12.2 °, 24.7 ° and 28.2 °corresponded to the Silk I structure, and those at 9.1 ° and 20.7 ° related to Silk II. The XRD curves of films were fitted to analyze their crystallinity. The crystallinities of traditional and improved silk films were 25.53% and 26.96% (Fig. 6c). The modified film has higher crystallinity and more β-sheet than the traditional film. The long molecular chains of the improved films are more likely to form crystallization zones under stronger molecular interactions, resulting in increased crystallinity. The results are in good agreement with those of Fourier transform infrared spectroscopy.

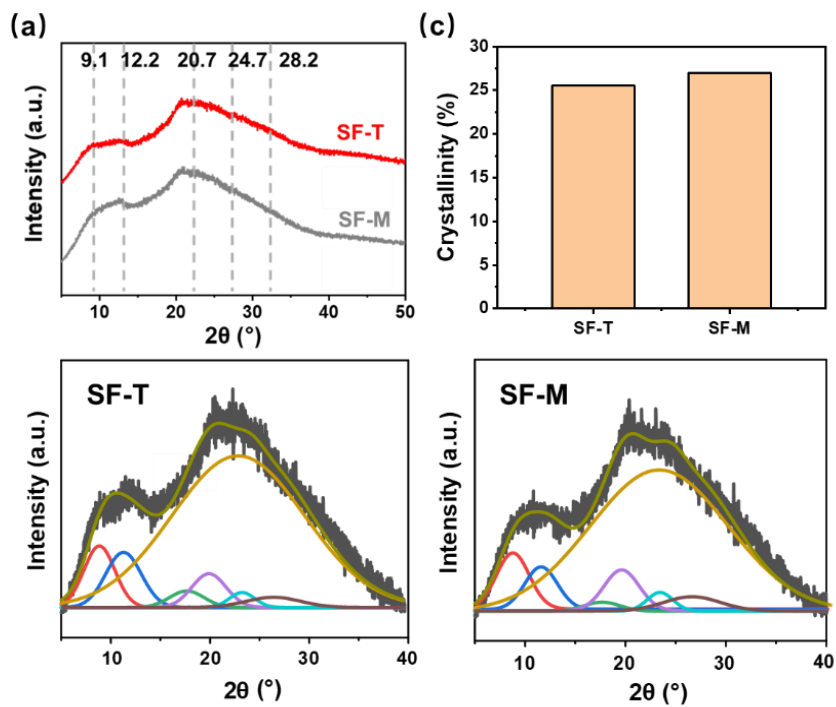

Fig.6 (a) XRD curves, (b) XRD pattern peak fitting diagram, (c) Crystallinity.

3.2.3 Mechanical properties

From Figure 7a, the tensile strength of SF-T and SF-M were 72.62 MPa and 87.91 MPa in the dry state. The modified silk film has higher tensile properties. Due to the higher β-sheet structure, the improved film has stronger intermolecular interaction and greater tensile strength than the traditional silk. However, the dry tensile elongation of the two films is only about 3%. It is shown that the silk film not only has high tensile strength, but also has the characteristics of hard and low flexibility, which has become the reason for limiting the application of silk film.

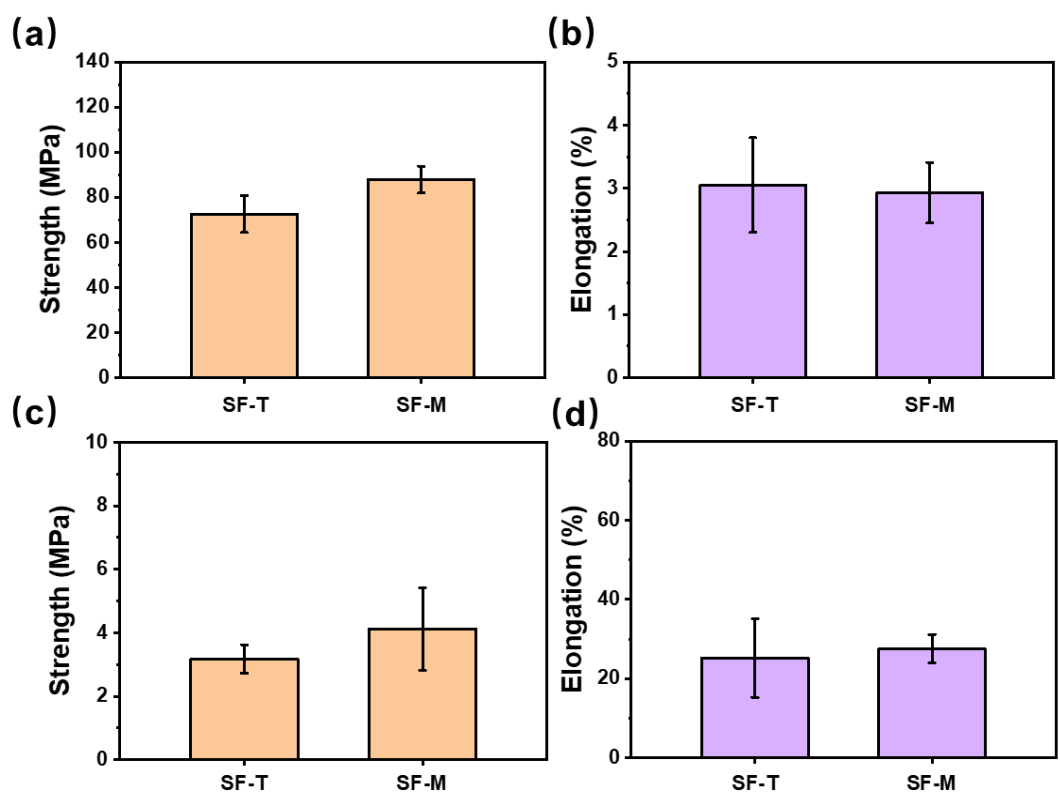

Fig.7 Mechanical properties of silk films (a) Dry tensile strength, (b) Dry tensile elongation, (c) Wet tensile strength, (d) Wet tensile elongation.

In the wet state, the tensile strength of films decreased obviously. Two wet tensile strengths of traditional and modified silk films were 3.18 MPa and 4.12 MPa. SF-M with more β-sheets have stronger wet strength. Water molecules entered into the silk fibroin in case of destroy the original hydrogen bond network of the protein molecules and the β-sheet structure was partially transformed into random coil, result in the wet tensile strengths were greatly reduced. The wet tensile elongations of two kinds of films were 25.22% and 27.57%. There is not much difference between the two silk films. Water within and between silk fibroin could increase the flexibility of the films. The greater the molecular weight and the longer the molecular chain of silk films, the higher the ductility of the molecular chain and the wet tensile elongation.

3.2.4 Air permeability, hydrophilicity and swelling performance

The water vapor transmittance of traditional and modified films were 0.069 g·cm$^{-2}$·24 h and 0.068 g·cm$^{-2}$·24 h (Fig. 8a). The permeabilities of two films were both low. The water vapor transmittance of SF-M is lower than that of SF-T with the increase of molecular weight of silk fibroin, the permeabilities of silk films decreased slightly and the β-sheet in silk increased. Therefore, The β-sheet structures were relatively more, and water vapor was more difficult to penetrate the films in that the modified silk films had lower water vapor permeability.

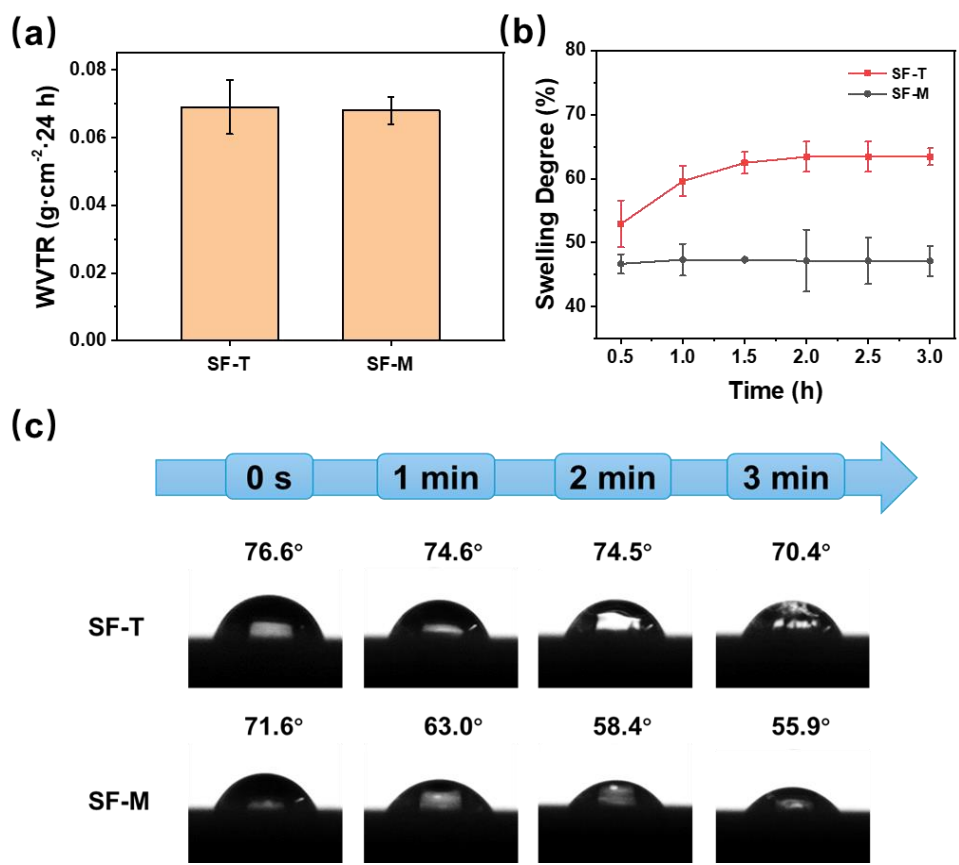

Fig.8 (a) Water vapor transmittance ratio, (b) Swelling degree, (c) Contact angle.

Between 0.5 and 2 hours, the SF-M samples showed varying degrees of swelling: 46.69%, 47.33%, 47.34% and 47.17% (Fig. 8b). The swelling capacity strengthened when the SF-T's swelling degree reached 62.47% after 1.5 hours. Water molecules found it more difficult to penetrate the modified silk molecules due to their denser structure and reduced porosity. The degree of swelling was reduced in the modified silk films. The two films stabilized after two hours. Since there was little interaction between the SF-T protein molecular chains, water molecules were able to more easily pass through the molecular chain gap, increasing the degree of swelling and fortifying the ability to absorb liquids.

The contact angles of the films were 76.6 ° and 71.6 ° at 0 s with the traditional and modified films, indicating that the films were hydrophilic (Fig. 8c). The changes of the contact angle were not obvious, it means that the silk surfaces were difficult to absorb solution. The α-helix and random coil structure were larger in two films. It was richer in hydrophilic amino acids, which enhanced the hydrophilicity of protein. At the same time, the increase of β-sheet structure made the silk films less absorbent after water vapor treatment.

3.2.5 Degradation performance

Silk fibroin films were placed into PBS buffer, PBS solution of 15 μg/mL and 100 μg/mL protease K, then their degradations were assessed according to weight loss (Fig. 9a). Silk films degradation rate in PBS solution was extremely slow. The residual masses percentage of SF-T and SF-M were about 95% within 16 days and the residual mass fraction of SF-T was slightly lower than that of SF-M because of the lower β-sheet content. There was almost no degradation of the two films within two weeks. The content of β-sheet increased after water vapor crosslinking made them had high stability in PBS solution.

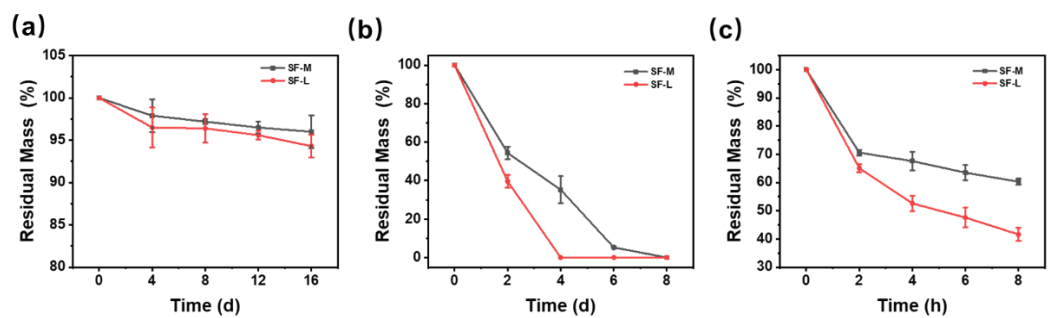

Fig.9 Degradation behavior of silk films (a) In PBS solution, (b) In PBS solution of 15 μg/mL protease K, (c) In PBS solution of 100 μg/mL protease K.

Protease K was used to accelerate the degradation of silk films. The protein films were degraded in PBS solution of protease K at 15 μg/mL and the residual mass fractions of SF-T and SF-M fibroin protein films were found to be 39.60% and 54.36% on the second day and the residual mass fraction of SF-T was found to be completely degraded on the fourth day, the residual mass fraction of SF-M was 35.24% at the same time (Fig. 9b). On the eighth day, all silk films were degraded. The results showed that SF-T with the more β-sheet structure were more stable and degraded slowly.

The higher the concentration of protease K, the faster the degradation of silk films. The films were degraded in PBS degradation solution with protease K concentration of 100 μg/mL. SF-T was found to have a faster degradation rate than SF-M, consistent with the previous conclusion that its lower β-sheet content led to faster degradation (Fig. 9c).

## 4 Conclusions

In conclusion, the modified degumming method by exposing the modified degummed silk fibers to 0.05% sodium carbonate solution for 0.5 hours and pure water for 1.0 hours was superior to the traditional method. the modified silk fiber average diameters (12.06 μm) were thicker than the traditional one(11.74 μm). And their β-sheet content (53.43% vs 46.68%), the crystallinity (44.92% vs 42.72%). The dissolved modified regenerated silk fibroin's molecular weights were primarily within the 100 ~ 300 kDa range rather than 72 ~ 300 kDa of traditional method. In addition to effectively eliminating sericin, this condition preserved the higher molecular weight of silk fibroin while slowing down its degradation in comparison to conventionally degummed silk.

The modified silk films exhibited higher β-sheet content (35.22% vs 32.82%), crystallinity (26.96% vs 25.53%), and dry tensile strength (87.91 MPa vs72.62 MPa) in comparison to the traditional silk films. On the other hand, the modified films with the slower degradation rate had lower swelling and water vapor permeability and higher molecular weight.

To put it briefly, the improved degumming scheme not only can clean sericin and retain higher molecular weight, but also has better film properties than those prepared by traditional methods. This degumming method enhanced silk films outperform and lays a foundation for the development of high-performance silk materials.